# Structure factor for decorated Penrose tiling

Bartlomiej Kozakowski, Janusz Wolny

Faculty of Physics and Applied Computer Science,
AGH University of Science and Technology,
30-059 Kraków, al. Mickiewicza 30, POLAND

**Abstract**
The structure factor for arbitrary decorated Penrose tiling has been calculated in average unit cell description. Analytical expression for the structure factor has been derived in physical space. The obtained formulas can be straightforward extended to some imperfect structures, including phasons or phonons and also some other defects.

**Introduction**

The best known model of 2D quasicrystals is the Penrose tiling. Similar tilings have been found in many experimental works for decagonal quasicrystals [1-3]. Aperiodic tiling of the plain built by two elements (kite and dart) was discovered by Penrose in 1974 [4]. Later de Brujin [5] used the canonical projection method to derive the fundamental properties of Penrose tilings. Kramer and Neri [6] and Levine and Steinhardt [7] independently developed a three-dimensional generalization of Penrose tilings by projecting a three dimensional cut through a higher-dimensional periodic lattice. Usually the quasicrystalline structure is described in higher-dimensions using the "cut and project method" discussed in many papers [5-16]. For Penrose tiling one needs to operate in 5D space with 2D physical space and 3D inner space (perp-space). The so called atomic surface consists of two small and two big pentagons and also two additional points. The structure factor calculations can be easily accomplished in perp-space. However, there is a serious problem with including additional decorating atoms of the rhombuses. The problem even increases for defected structures and only phasons have been successfully included. Influence of phonons and other defects on the structure factor is rather impossible to be considered in higher-dimensional analysis.

In this paper we have calculated the structure factor in physical space using the average unit cell approach, which is based on a concept of a reference lattice (reference grid) [17,18]. Let's suppose we have a 1 dimensional set $\{r_n\}$. Its points represent atoms whose scattering intensities are equal to unity. For a given scattering vector $k$ one can construct a lattice with the lattice constant $\lambda = 2\pi/k$. The series of relative positions:

$$u_n \equiv r_n \bmod(\lambda) \qquad (1)$$

are called the displacements sequence of $r_n$ (induced by the reference lattice with periodicity $\lambda$). The structure factor can then be written as:

$$F_N(k) = \frac{1}{N}\sum_{n=1}^{N} \exp(ikr_n) = \frac{1}{N}\sum_{n=1}^{N} \exp(iku_n) \qquad (2)$$

In the limit $N \to \infty$, for special sets such as the set of vertices of the Penrose tiling (more generally, of ideal quasicrystals), and for modulated structures [17], a function $P(u)$ exists such that

$$F(k) = \lim_{N \to \infty} (F_N(k)) = \int_0^\lambda P(u) \exp(iku) \, du \tag{3}$$

where $P(u)$ is the probability distribution of distances $u_n$, and we call it an average unit cell for particular $k$ vector.

For all other structures (i.e. random, amorphous) where the diffraction peak intensities scale not as $N^2$ (not Bragg reflections), the average unit cell can also exist and approximates the structure by following:

$$F_N(k) \approx \int_0^\lambda \overline{P_N(u)} \exp(iku) \, du \tag{4}$$

where $\overline{P_N(u)}$ is an average distribution of atomic distances to the reference grid over statistical ensemble of clusters. For big enough $N$ expression (4) usually approximates the structure factor better than it could be determined for a given experimental accuracy. There are also structures (like the Thue-Morse structure) where some particular subsets of numbers of atoms should be considered for the structure factor calculation.

To describe the diffraction pattern of 1D modulated structure one needs two vectors: $k$ and $q$, i.e. the main scattering vector $k$ (describing the main diffraction peaks) and the modulation vector $q$ (for the satellite reflections) [19]. This leads to the probability distribution $P(u,v)$, where $u$ and $v$ are the shortest distances of the atomic position from the appropriate points of two reference lattices, for $k$ and $q$ vectors respectively. For 2D modulated structures four vectors are required: $\mathbf{k}_i$, $\mathbf{q}_i$, $i=1,2$, and the corresponding probability distribution is $P(u_1, u_2, v_1, v_2)$. The displacements $u_1$, $u_2$ are associated with the main wave vectors $\mathbf{k}_1$, $\mathbf{k}_2$, and the displacements $v_1$, $v_2$ – with the modulation vectors $\mathbf{q}_1$, $\mathbf{q}_2$.

It was shown [18] that the average unit cell is an oblique projection of the atomic surface in higher-dimensions. The perp-space atomic surface is projected orthogonally to the chosen scattering vector written in higher-dimensions (see Fig. 1 in [18]). In that sense, the atomic surface is nothing else but the average unit cell risen up to higher-dimensions. Both analysis, i.e. the higher-dimensional cut and projection analysis and the average unit cell are based on the same statistical approach. In previous papers [18,20] using an average unit cell approach the structure factor for Penrose tiling without any decorations was derived. In the present paper the problem for decorated structures have been solved by using more accurate algorithm for obtaining the structure factor.

When it is possible we also use the higher-dimensions to speed up the calculations, but in principle all the calculations as well as the final result can be fully accomplished in physical space only. The obtained analytical expression for the structure factor can then be easily extended to some other defects, including phonons and phasons.

**The average unit cell for Penrose tiling.**

In this chapter the average unit cell is calculated for the Penrose tiling. Such approach allows obtaining analytical expression for the structure factor. The following vector notations are used: $\mathbf{k} = (k_x, k_y)$ is a scattering vector and its Cartesian coordinates in physical space; $\mathbf{K}=[k_1, k_2, k_3, k_4, k_5]$ is a 5D reciprocal lattice vector; $\mathbf{K}=\{k_x, k_y, k_{\perp x}, k_{\perp y}, k_z\}$ is the 5D scattering vector written for physical and internal (perpendicular) spaces.

The Penrose tiling consists of two rhombuses: the thick one (*L*) and the thin one (*S*). We define the reference frame as it is shown in Fig. 1. Most of our calculations will be illustrated for the thick rhombus shown in Fig. 1 and denoted by indices 23, which means that the bounds are oriented along the "2" [0,1,0,0,0] and "3" [0,0,1,0,0] directions (in square brackets the 5D coordinates are given). Similarly for the thin rhombus, the one with indices $1\bar{4}$ has been chosen as an example.

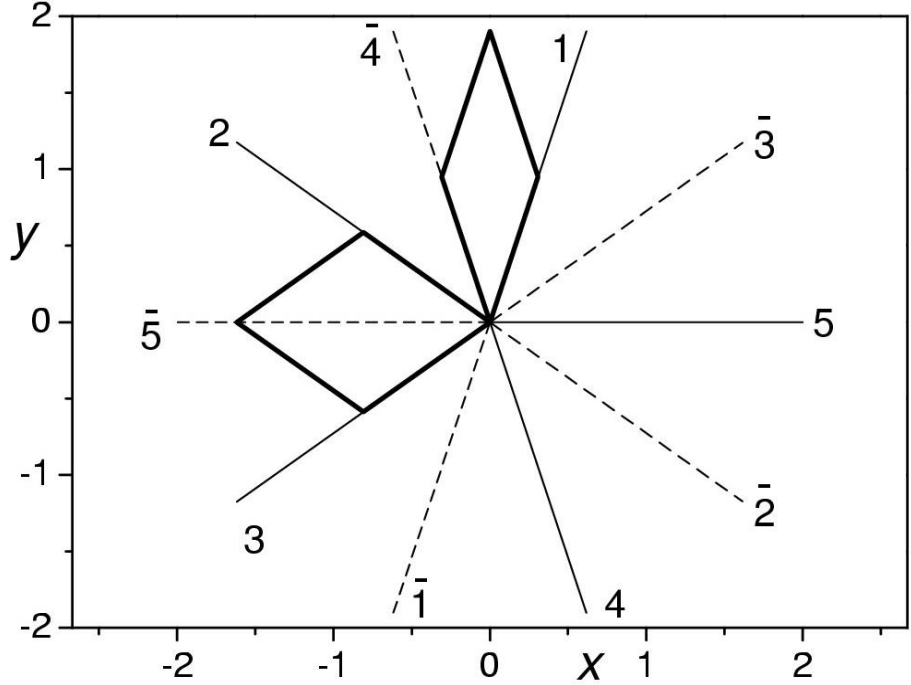

***Fig.1***. *Setting of directions for Penrose tiling. Two selected rhombuses, $S_{1\bar{4}}$ and $L_{23}$, used as the basic ones, have been also drown.*

Fig. 2 shows the way to find part of the atomic surface at $z=1$ for a point placed at the origin and belonging to $L_{23}$ rhombus. The neighbouring vertices are at: [0,1,0,0,0] and [0,0,1,0,0], and they are placed at $z=2$. In a perp-space the two bounds are given by two vectors: $\left(\frac{1}{2\tau},\pm\frac{1}{2}\sqrt{\tau+2},1\right)$, where $\tau \equiv \frac{1+\sqrt{5}}{2} \approx 1.618$, shown in the figure as **AA₁** and **AA₂**. The required distribution is then the triangle ABC, which is a common part of the small pentagon ($z=1$) and the big ones ($z=2$) when shifted by the two vectors $\mathbf{w_1}=\mathbf{A_1A}$ and $\mathbf{w_2}=\mathbf{A_2A}$ appropriately.

For the atoms placed at the corners of the rhombus the average unit cell consists of two triangles as it is shown in Figs 3&4 for $S_{1\bar{4}}$ and $L_{23}$ rhombuses respectively. The corresponding perp-space coordinates are then given in Table 1.

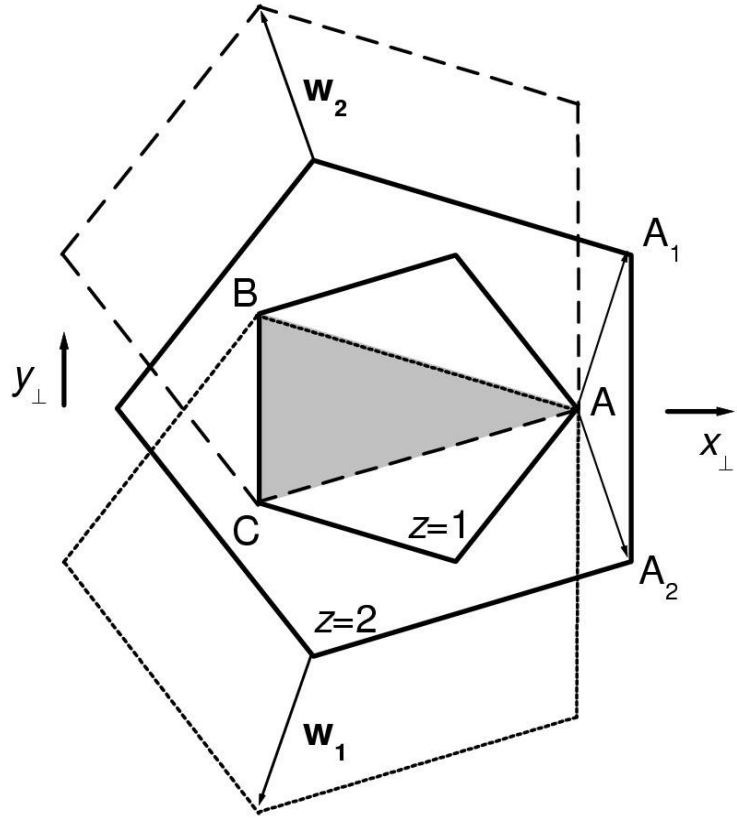

***Fig. 2***. *This figure shows the way to construct the triangular distributions in per-space for $L_{23}$ rhombus at z=1 (see also Table 1).*

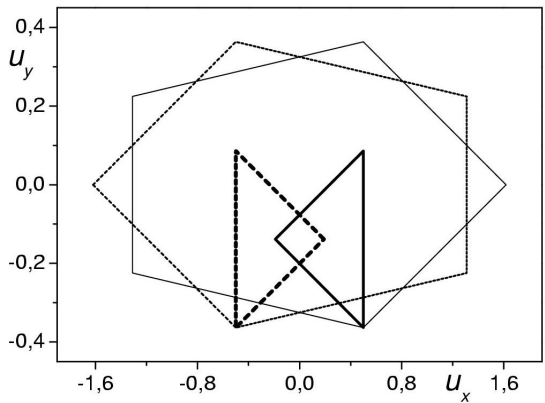

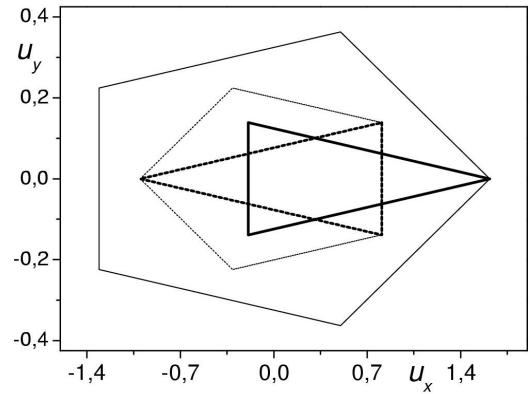

***Fig.3***. *Statistical distribution of position (0,0) of the thin rhombus in the average unit cell. The solid and broken lines are for z=2 and 3 respectively. The thin lines (solid and broken) are the contours of the pentagonal atomic surfaces when projected to the physical space. The thick lines (solid and broken) are the triangular distributions obtained for the small rhombus shown in Fig. 1.*

***Fig.4.*** *The same distribution as in Fig. 2 but for thick rhombus and for z=1 (broken lines) and z=2 (solid lines).*

Table 1: Vertices of triangular distributions being a part of the atomic surface and corresponding to two rhombuses shown in Fig. 1.

| $L_{23}$ | | | | $S_{1\bar{4}}$ | | | |
|---|---|---|---|---|---|---|---|
| $z=1$ | | $z=2$ | | $z=2$ | | $z=3$ | |
| $x_{\perp L}$ | $y_{\perp L}$ | $x_{\perp L}$ | $y_{\perp L}$ | $x_{\perp S}$ | $Y_{\perp S}$ | $x_{\perp S}$ | $y_{\perp S}$ |
| $2\tau\cdot c_1$ | 0 | $-2(\tau+1)\cdot c_1$ | 0 | $(\tau-1)\cdot c_1$ | $-(\tau-1)\cdot s_1$ | $-(\tau-1)\cdot c_1$ | $-(\tau-1)\cdot s_1$ |
| $-(\tau+1)\cdot c_1$ | $(\tau-1)\cdot s_1$ | $(\tau-1)\cdot c_1$ | $(\tau-1)\cdot s_1$ | $-\tau\cdot c_1$ | $(2-\tau)\cdot s_1$ | $\tau\cdot c_1$ | $(2-\tau)\cdot s_1$ |
| $-(\tau+1)\cdot c_1$ | $-(\tau-1)\cdot s_1$ | $(\tau-1)\cdot c_1$ | $-(\tau-1)\cdot s_1$ | $-\tau\cdot c_1$ | $-\tau\cdot s_1$ | $\tau\cdot c_1$ | $-\tau\cdot s_1$ |

where:

$$c_j = \cos\left(\frac{2\pi j}{5}\right); \quad s_j = \sin\left(\frac{2\pi j}{5}\right); \quad \mathbf{C}_j \equiv \begin{pmatrix} c_j & s_j \\ -s_j & c_j \end{pmatrix}$$

For both rhombuses (*L* and *S*) there are two symmetrical triangular distributions. One distribution corresponds to particular orientation of rhombus, the other one – to the same rhombus but rotated by 180 degree. These two orientations are symmetrically related and they are not distinguished for simple Penrose tiling. However, for decorated tilings these two orientations could be different. For the simplicity of further calculations one triangular distribution for each rhombus is used, i.e. at $z=1$ for *L* and at $z=2$ for *S*. The mirror symmetry of the distribution is automatically considered by taking into account only the real part of the Fourier transform and neglecting the imaginary part. For symmetrical distributions the imaginary part of the Fourier transform is equal to zero. The used direct relation between the particular rhombus in a given orientation and the appropriate triangular distribution, while considering only the real part of the Fourier transform leads to essential simplification of the derived formulas without any restriction to the correctness of derived formulas for the structure factor.

The discussed triangles lie on the atomic surface with different values of *z* component. Components of triangular distributions for the other rhombuses ($t \in \{L, S\}$) can be obtained using the rotation:

$$\begin{pmatrix} x_{\perp tj} \\ y_{\perp tj} \end{pmatrix} = \mathbf{C}_{2j} \begin{pmatrix} x_{\perp t} \\ y_{\perp t} \end{pmatrix} \quad (5)$$

Index *j* distinguishes between different rhombuses as it is given in the Table 2.

| j | L | S |
|---|---|---|
| 0 | 23 | $1\bar{4}$ |
| 1 | 12 | $5\bar{3}$ |
| 2 | 51 | $4\bar{2}$ |
| 3 | 45 | $3\bar{1}$ |
| 4 | 34 | $2\bar{5}$ |

Table 2

Now the triangular distributions lying on the atomic surface will be projected onto physical space. The oblique projection is perpendicular to the two 5D scattering vectors given by $\mathbf{K_1}=[0,0,\bar{1},\bar{1},0]$; $\mathbf{K_2}=[\bar{1},\bar{1},0,0,0]$ and to the $\tau$-times shorter two modulation vectors $\mathbf{Q_1}=[1,0,0,0,0]$; $\mathbf{Q_2}=[0,0,0,1,0]$, respectively, what leads to the probability distribution $P(u_1, u_2, v_1, v_2)$. The four vectors: $\mathbf{K}_i$ and $\mathbf{Q}_i$, $i=1,2$, are chosen in such way that their real-space components, $\mathbf{k}_i$ and $\mathbf{q}_i$, $i=1,2$ fully describe the analysed diffraction pattern by four indices $(n_1, n_2, m_1, m_2)$. In the used notation $n_1$ and $n_2$ index the positions of the main reflections and $m_1, m_2$ – their satellites. It is obvious that some other sets of vectors can be also used.

The physical and internal reciprocal spaces are spanned on vectors:

$$\hat{\mathbf{K}}_x = \frac{4\pi}{5}[c_1,c_2,c_3,c_4,c_5] = \frac{4\pi}{5}[c_1,c_2,c_2,c_1,1]$$

$$\hat{\mathbf{K}}_y = \frac{4\pi}{5}[s_1,s_2,s_3,s_4,s_5] = \frac{4\pi}{5}[s_1,s_2,-s_2,-s_1,0]$$

$$\hat{\mathbf{K}}_{\perp x} = \frac{4\pi}{5}[c_2,c_4,c_6,c_8,c_{10}] = \frac{4\pi}{5}[c_2,c_1,c_1,c_2,1] \quad (6)$$

$$\hat{\mathbf{K}}_{\perp y} = \frac{4\pi}{5}[s_2,s_4,s_6,s_8,s_{10}] = \frac{4\pi}{5}[s_2,-s_1,s_1,-s_2,0]$$

$$\hat{\mathbf{K}}_z = \frac{2\pi}{5}[1,1,1,1,1]$$

When projected onto physical and internal spaces one gets for $\mathbf{K_1}, \mathbf{K_2}, \mathbf{Q_1}, \mathbf{Q_2}$:

$$\mathbf{K}_1 = \frac{4\pi}{5}\{c_1\tau, s_1\tau, c_2(\tau-1), s_2(\tau-1), -1\}$$

$$\mathbf{K}_2 = \frac{4\pi}{5}\{c_1\tau, -s_1\tau, c_2(\tau-1), -s_2(\tau-1), -1\}$$

$$\mathbf{Q}_1 = \frac{4\pi}{5}\left\{c_1, s_1, c_2, s_2, \frac{1}{2}\right\} \quad (7)$$

$$\mathbf{Q}_2 = \frac{4\pi}{5}\left\{c_1, -s_1, c_2, -s_2, \frac{1}{2}\right\}$$

The physical space reciprocal vectors, $\mathbf{k}_i$, $\mathbf{q}_i$, $i=1,2$, are formed by the first two coordinates of $\mathbf{K}_i$, $\mathbf{Q}_i$, and the perp-space vectors, $\mathbf{k}_{i\perp}$, $\mathbf{q}_{i\perp}$, by the last three coordinates.

The lengths of the reciprocal vectors in physical and perp-spaces are

$$|\mathbf{k_1}| = |\mathbf{k_2}| = \frac{4\pi}{5}\tau \equiv k_0 \approx 4.067; \qquad |\mathbf{k}_{1\perp}| = |\mathbf{k}_{2\perp}| = k_0(2-\tau) \approx 1.553$$

$$|\mathbf{q_1}| = |\mathbf{q_2}| = |\mathbf{q}_{1\perp}| = |\mathbf{q}_{2\perp}| = \frac{4\pi}{5} \equiv \frac{k_0}{\tau} \approx 2.513 \quad (8)$$

Vectors **k₁** and **q₁** are directed at the angle of 72° to the *x* axis (direction 1 in Fig. 1) and vectors **k₂** and **q₂** are directed at the angle -72° to the axis of *x* (direction 4 in Fig. 1). An arbitrary diffraction peak for scattering vector **k** can be then expressed as a linear combination of $\mathbf{k}_i$ and $\mathbf{q}_i$ (*i*=1,2) with indices $n_i$ and $m_i$ appropriately, and its components ($k_x$,$k_y$) are equal to:

$$k_x = k_0 c_1 \left( n_x + \frac{m_x}{\tau} \right) \quad (9)$$

$$k_y = k_0 s_1 \left( n_y + \frac{m_y}{\tau} \right) \quad (10)$$

where: $n_x \equiv n_1 + n_2, \quad m_x \equiv m_1 + m_2, \quad n_y \equiv n_1 - n_2, \quad m_y \equiv m_1 - m_2.$ (11)

For the scattering vector given above one can calculate the structure factor in an average unit cell approach as following [17-19]:

$$F = \iiint\!\!\int P(u_1, u_2, v_1, v_2) \exp\left( ik_0 \left( n_1 u_1 + n_2 u_2 + \frac{1}{\tau}(m_1 v_1 + m_2 v_2) \right) \right) du_1 du_2 dv_1 dv_2 \quad (12)$$

where $P(u_1, u_2, v_1, v_2)$ is a probability distribution which defines an average unit cell for the structure. For the Cartesian coordinates one gets: ($u_x$, $u_y$) and ($v_x$, $v_y$). The components ($u_1$, $u_2$) along $\mathbf{k}_1$ and $\mathbf{k}_2$, and ($v_1$, $v_2$) along $\mathbf{q}_1$ and $\mathbf{q}_2$, are given then by the orthogonal projections of the above vector onto the two directions: $\hat{\mathbf{1}} = (c_1, s_1)$ and $\hat{\mathbf{2}} = (c_1, -s_1)$, which leads to the following

$$u_1 = c_1 u_x + s_1 u_y; \quad v_1 = c_1 v_x + s_1 v_y$$

$$u_2 = c_1 u_x - s_1 u_y; \quad v_2 = c_1 v_x - s_1 v_y \quad (13)$$

Then one can write:

$$F = \iiint\!\!\int P(u_x, u_y, v_x, v_y) \exp\left[ ik_0 \left( c_1 \left( n_x u_x + m_x \frac{v_x}{\tau} \right) + s_1 \left( n_y u_y + m_y \frac{v_y}{\tau} \right) \right) \right] du_x du_y dv_x dv_y \quad (14)$$

Formula (14) is the same as (12) but written for the Cartesian coordinates, which has very practical meaning.

Diffraction peaks are placed at reciprocal lattice points given by a product of 5D vectors: **K·R** = 2π*n*. The multiplicity of 2π does not count for the structure factor and one can put zero on the right hand of the above equation.
The perp-space coordinates $x_\perp$, $y_\perp$, $z$ of vertices of the triangular distributions are given in Table 1. For vectors $\mathbf{k}_i$, *i*=1,2, one gets (modulo 2π on the right hand of the equation):

$$k_{ix} u_x + k_{iy} u_y + k_{i\perp x} x_\perp + k_{i\perp y} y_\perp + k_{iz} z = 0 \quad (15)$$

According to (5) and (6) one can also write:

$$k_{0x} \equiv k_{1x} = k_{2x} = c_1 k_0 \approx 1.257$$
$$k_{0y} \equiv k_{1y} = -k_{2y} = s_1 k_0 \approx 3.868$$
$$k_{0\perp x} \equiv k_{1\perp x} = k_{2\perp x} = k_{0x} \approx 1.257 \quad (16)$$
$$k_{0\perp y} \equiv k_{1\perp y} = -k_{2\perp y} = -k_{0y}(2\tau - 3) \approx -0.913$$
$$k_{0z} \equiv k_{1z} = k_{2z} = -(\tau - 1)k_0 \approx -2.513$$

Equations (15) and (16) lead to:

$$u'_x \equiv u_x + \frac{k_{0z}}{k_{0x}} z = -\frac{k_{0\perp x}}{k_{0x}} x_\perp = -x_\perp$$
$$u_y = -\frac{k_{0\perp y}}{k_{0y}} y_\perp = \frac{1}{\tau^3} y_\perp \quad (17)$$

or in a matrix representation: $\quad \mathbf{u} = \mathbf{A}\mathbf{r}_\perp; \quad \mathbf{r}_\perp = \mathbf{A}^{-1}\mathbf{u} \quad (18)$

where:

$$\mathbf{u} \equiv \begin{pmatrix} u'_x \\ u_y \end{pmatrix} = \begin{pmatrix} u_x - 2z \\ u_y \end{pmatrix}; \quad \mathbf{r}_\perp \equiv \begin{pmatrix} x_\perp \\ y_\perp \end{pmatrix}; \quad (19)$$

$$\mathbf{A} \equiv \begin{pmatrix} -\dfrac{k_{0\perp x}}{k_{0x}} & 0 \\ 0 & -\dfrac{k_{0\perp y}}{k_{0y}} \end{pmatrix} = \begin{pmatrix} -1 & 0 \\ 0 & \tau^{-3} \end{pmatrix}; \quad \mathbf{A}^{-1} = \begin{pmatrix} -1 & 0 \\ 0 & \tau^3 \end{pmatrix} \quad (20)$$

Similar calculations performed for modulation vectors $\mathbf{q}_1$ and $\mathbf{q}_2$ give:

$$v'_x \equiv v_x + \frac{q_{0z}}{q_{0x}} z = -\frac{q_{0\perp x}}{q_{0x}} x_\perp = \tau^2 x_\perp$$
$$v_y = -\frac{q_{0\perp y}}{q_{0y}} y_\perp = -\frac{1}{\tau} y_\perp \quad (21)$$

From (17-21) it follows that:
$$\mathbf{v} = -\tau^2 \mathbf{u} \quad (22)$$

where:
$$\mathbf{v} \equiv \begin{pmatrix} v'_x \\ v_y \end{pmatrix} = \begin{pmatrix} v_x + \tau z \\ v_y \end{pmatrix} \quad (23)$$

The above also means that the probability distribution is nonzero only along the line given by (22).

**Decorated structures**

Now let's suppose that the thick rhombus (*L*) is decorated by an atom placed at position $\mathbf{r}_L$. Similarly, the thin rhombus (*S*) is decorated by an atom placed at position $\mathbf{r}_S$. Then, using (14) and (22), the structure factor for a given type of rhombus $t \in \{S, L\}$ is:

$$F_t = \sum_{j=0,4} F_{tj} \exp(i\mathbf{k} \cdot \mathbf{r}_{tj}) \qquad (24)$$

where:

$$\mathbf{r}_{tj} = \mathbf{C}_j \mathbf{r}_t \qquad (25)$$

$$F_{tj} = \exp(i\varphi_z) \iint_{\Delta_{tzj}} \exp(i\boldsymbol{\kappa} \cdot \mathbf{u}) d^2 u \qquad (26)$$

$$\varphi_z = z(\tau - 1)(k_x + m_x(3-\tau)k_0) \qquad (27)$$

$$\boldsymbol{\kappa} = \begin{pmatrix} k_x - m_x(2\tau-1)k_{0x} \\ k_y - m_y(2\tau-1)k_{0y} \end{pmatrix} \qquad (28)$$

The integration is over the area of triangles $\Delta_{tzj}$ which vertices can be obtained from the following relation:

$$\mathbf{u}_{tj} = \mathbf{A}^{-1}\mathbf{C}_{2j}\mathbf{A}\mathbf{u}_t = \begin{pmatrix} c_{2j} & -\tau^3 s_{2j} \\ \tau^{-3} s_{2j} & c_{2j} \end{pmatrix} \cdot \begin{pmatrix} u'_{tx} \\ u_{ty} \end{pmatrix} \qquad (29)$$

or in a more practical way:

$$\mathbf{u}_{tj} = \mathbf{A}^{-1}\mathbf{C}_{2j}\mathbf{r}_{t\perp} = \begin{pmatrix} -c_{2j} & -s_{2j} \\ -\tau^{-3} s_{2j} & \tau^{-3} c_{2j} \end{pmatrix} \cdot \begin{pmatrix} x_{t\perp} \\ y_{t\perp} \end{pmatrix} \qquad (30)$$

where the perp-space components are given in Table 1.
Finally, for $n_L$ atoms decorating the thick rhombus (at the positions $\mathbf{r}_{Ln}$) and $n_S$ atoms decorating the thin one (at the positions $\mathbf{r}_{Sn}$), one obtains the analytical expression for the structure factor of decorated Penrose tiling:

$$F = \mathrm{Re}\left\{ \sum_{j=0}^{4} \left( B_{Lj}(\mathbf{k}) \sum_{n=1}^{n_L} \exp(i\mathbf{k} \cdot \mathbf{r}_{Lnj}) + B_{Sj}(\mathbf{k}) \sum_{n=1}^{n_S} \exp(i\mathbf{k} \cdot \mathbf{r}_{Snj}) \right) \right\} \qquad (31)$$

where

$$B_{Lj}(\mathbf{k}) = \Gamma_{L1j}(\boldsymbol{\kappa}) \exp(i\varphi_1) \qquad (32)$$

$$B_{Sj}(\mathbf{k}) = \Gamma_{S2j}(\boldsymbol{\kappa}) \exp(i2\varphi_1) \qquad (33)$$

$$\varphi_1 = \varphi_{z=1} = (\tau-1)(k_x + m_x(3-\tau)k_0) \qquad (34)$$

Values of $\Gamma_{tzj}$ are the Fourier transforms of triangular distributions given by (30) and they are discussed in the Appendix. All the values of $B_{tj}(\mathbf{k})$ in (31) do not depend on decoration and need to be calculated for the Penrose tiling only once and then remembered as an appropriate array of numbers (one set of values for each diffraction peak). This property is extremely important for any refinement of the structure. Expression for the structure factor of arbitrary decorated Penrose tiling is not more complicated than for ordinary crystals. The only difference is that there are two decorated elements (thick and thin rhombuses) which should be considered separately. Additionally, a given rhombus appears in five different orientations which is automatically included in the sum over $j=0,4$. One can also extend the formula (31) to some crystals imperfections, like phonons or phasons. Phonons in the first approximation lead to the well-known Debye-Waller factor which can be easily included in (31). Phasons change only the probability distribution and modification of (31) is then straightforward. One has to use appropriately modified probabilities of sites' occupation.

*Example 1. Ordinary Penrose tiling.*

In the first example (Fig. 5a) let's consider an ordinary Penrose tiling where identical atoms (with the atomic form-factor equals to unity) are placed at the corners of rhombuses. In this case four atoms decorate the thick and thin rhombuses. Each corner of the rhombus is decorated by one atom with the probability proportional to the inner angle of the vertex, which gives the values: {0.3, 0.2, 0.3, 0.2} for the thick rhombus and {0.1, 0.4, 0.1, 0.4} for the thin one. In Figs 6&7 two cross-sections of the obtained diffraction pattern are shown. The agreement between the intensities calculated for a big enough cluster and by (31) is perfect.

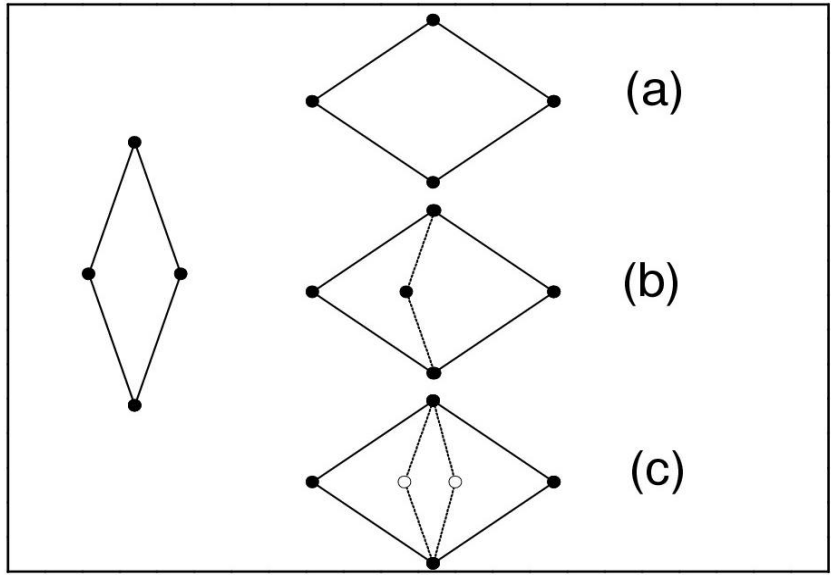

*Fig. 5*. *Three types of decorations used for the diffraction patterns analysis. In the case c) the atoms occupy the inner positions randomly.*

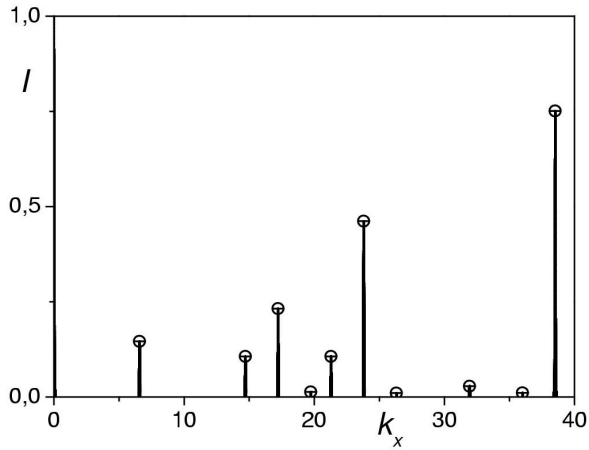
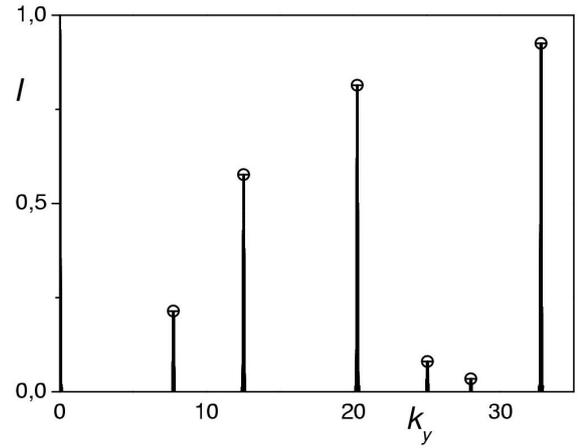

*Fig. 6.* The thick solid line represents the diffraction pattern along the x-direction for Penrose tiling calculated for a cluster of about 2400 atoms. The open circles represent the result obtained from equation (31) obtained for the structure factor.

*Fig. 7.* Similarly as in Fig. 5 but along the y-direction.

*Example 2. Decorated Penrose tiling*

In the second example the rhombuses are decorated as in *Example 1* (ordinary Penrose tilling) and additionally the thick rhombuses are decorated as it is shown in Fig. 5b. The diffraction pattern along *X*-direction is shown in Fig. 8. It consists of diffraction pattern calculated for a cluster of about 16000 atoms (full line) and intensities of the most intensive diffraction peaks calculated by obtained structure factor (31) – open circles. Full agreement between these calculations is observed. The diffraction pattern along *Y*-direction does not change compare to Fig. 7.

*Example 3. Disordered structure*

In this example it was supposed that the decorating atom from *Example 2* has two equivalent positions as it is shown in Fig. 5c. In this case the structure factor (31) can be also written:

$$F = \sum_{j=0}^{4} \left( B_{Lj}(\mathbf{k}) \sum_{n=1}^{n_L} p_{Ln} \exp(i\mathbf{k} \cdot \mathbf{r}_{Lnj}) + B_{Sj}(\mathbf{k}) \sum_{n=1}^{n_S} p_{Sn} \exp(i\mathbf{k} \cdot \mathbf{r}_{Snj}) \right) \qquad (35)$$

where $p_{tn}$ are the occupation probabilities. Supposing that these occupation probabilities for the flipping atom are equal to 0.5, one gets the diffraction pattern shown in Fig. 9. Also in this case a full agreement between direct calculations for big enough cluster of atoms and the structure factor obtained from (33) is observed. The super-structure reflections from Fig. 8 disappeared.

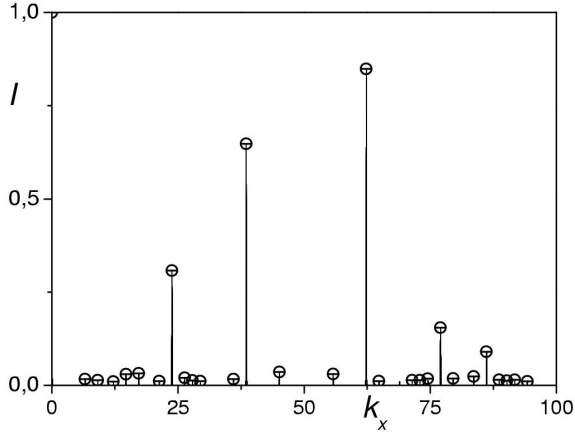 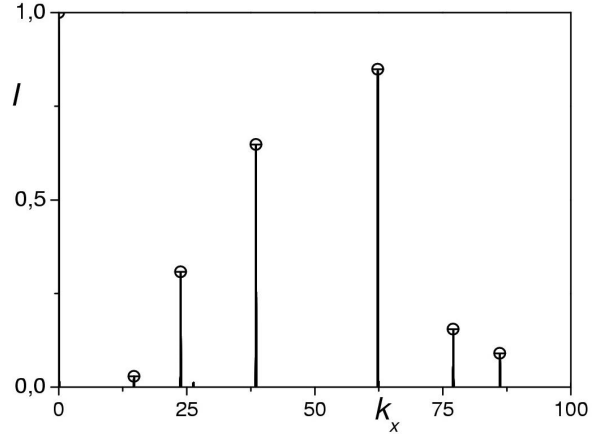

*Fig. 8*. *Diffraction patterns for the decorated Penrose tiling shown in Fig. 4b. Bars are for numerical calculations for a cluster of about 16000 atoms. Open circles are for the analytical expression given by (31). The compared results are fully equivalent.*

*Fig. 9.* *Diffraction patterns for the randomly decorated Penrose tiling shown in Fig. 4c. Bars are for numerical calculations for a cluster of about 16000 atoms. Open circles are for the analytical expression given by (31). The two results are fully equal at the positions of diffraction peaks. Compare to Fig. 8 many super-structure reflections were disappeared.*

**Conclusions**

Using the average unit cell approach the analytical expression for the structure factor of arbitrary decorated Penrose tiling was derived (similar calculations for 1D quasicrystals were presented in [21]). The obtained formula (31) is rather simple and very similar to the well-known expressions for ordinary crystals. It consists of phase factors ($\exp(i\mathbf{k}\cdot\mathbf{r})$) multiplied by the set of $B(k)$-values, which are the same for all possible decorations. This brings to the full parallelism between the structure factor for quasicrystals written in physical space and the one for ordinary crystals. During the refining procedure one has to adjust the phases (given by relative positions of the decorating atoms) keeping constant the multiplication factors, calculated only once for a given type of quasicrystal (in the discussed case – Penrose tiling).

The obtained formula for the structure factor is fully written in physical space, which is very useful for including some kind of imperfections, like phonons and phasons or some other defects. All known in classical crystallography approaches to these imperfections can be easily apply to quasicrystal's structure factor. Formula (31) is generally not applicable to random tilings as introduced by Elser [22]. For sufficiently weak forms of randomness, such as the example in the manuscript, formula (31) remains however valid. Stronger forms of randomness cannot be incorporated by easily extension of the present approach [see also 23].

Several decorated 2D quasicrystals have been tested and full agreement between analytical expression for the structure factor and the numerical calculations for the cluster of atoms have been observed. This means that the obtained formula for the structure factor is valid for arbitrary decoration of Penrose tiling. Although the 2D case have been analysed only, any extension to 3D case or even to arbitrary dimensions is straightforward.

**Appendix**

Fourier transform of triangular distributions:

$$\Gamma_{tzj}(\boldsymbol{\kappa}) \equiv \iint_{\Delta_{tzj}} \exp(i\boldsymbol{\kappa} \cdot \mathbf{u}) \mathrm{d}^2 u \tag{A1}$$

The distribution is non zero and constant inside the triangle $\Delta_{tzj}$. For the particular values of $t \in \{S,L\}$, $z=1,2$ and $j=0,1,2,3,4$, the vertices of the triangle are given by the following real space positions: $\mathbf{u}_p$, for $p=1,2,3$. The linear coefficients of the three lines which determine the triangle are: $a_{lm}$ for the line going through the vertices $\mathbf{u}_l$ and $\mathbf{u}_m$. Then the Fourier transform of the triangular distribution is given by the formula:

$$\Gamma_{tzj}(\boldsymbol{\kappa}) = \frac{1}{\kappa_y}\left(D_{12}(E_2 - E_1) + D_{23}(E_3 - E_2) + D_{31}(E_1 - E_3)\right) \tag{A2}$$

where:

$$D_{lm} = \frac{1}{\kappa_x + \kappa_y a_{lm}}; \qquad E_p = \exp(i\boldsymbol{\kappa} \cdot \mathbf{u}_p).$$